\documentclass[12pt]{article}

\usepackage{graphics}
\usepackage[dvips]{graphicx}

\begin{document}
{\footnotesize jcis@epacis.org}

\begin{center}

{\bf A Self-Consistent Extrapolation Method for the Complex Permittivity and Permeability Based on Finite Frequency Data}
\bigskip

{\small Christine C. Dantas$^a$ \footnote{E-mail Corresponding Author: christineccd@iae.cta.br},
Mirabel. C. Rezende$^{a,b}$ \\ and Simone S. Pinto$^a$}
\smallskip

{\small
$^a$ Materials Division (AMR), Instituto de Aeron\'autica e Espa\c co (IAE/DCTA), Pra\c ca Mal. Eduardo Gomes, 50, Vila das Ac\'acias, S\~ao Jos\'e dos Campos, SP, 12.228-904, Brazil\\

$^b$ Instituto Tecnol\'ogico de Aeron\'autica (ITA/DCTA), Pra\c ca Mal. Eduardo Gomes, 50, Vila das Ac\'acias, S\~ao Jos\'e dos Campos, SP, 12.228-900, S\~ao Jos\'e dos Campos, Brazil\\
}

{\footnotesize Received on *****, 2014 / accepted on *****, 2014}

\end{center}

\begin{abstract}
We describe a method of extrapolation based on a ``truncated'' Kramers-Kronig relation for the complex permittivity ($\epsilon$) and permeability ($\mu$) parameters of a material, based on finite frequency data. Considering a few assumptions, such as the behavior of the loss tangent and the overall nature of corrections, the method is robust within a small relative error, if the assumed hypotheses hold at the extrapolated frequency range.

\bigskip

{\footnotesize
{\bf Keywords}:  Permittivity, permeability, Kromers-Kronig relations, inverse problems, optimization, loss tangent.}
\end{abstract}

\bigskip
\bigskip

\textbf{1. INTRODUCTION}

\bigskip
\bigskip

The need for accurate measurements of the dielectric and magnetic
properties of materials, in order to improve the range and specificity of applications, is on increasing demand.  In particular, materials for electromagnetic applications are lossy and the measurement of their properties is challenging \cite{NIST}. Furthermore, the dielectric and magnetic properties of materials are only experimentally measured to within some finite frequency interval response. It is often the case that one is interested in expanding the experimental results throughout some other interval range, based on some theoretical assumptions, in order to predict/infer the behaviour of the material and hence expand its application horizon. 

The lossy properties of materials, more specifically, expressed in terms of the complex permittivity and permeability, can be understood by the following general arguments\footnote{The mathematical representation of the permittivity (or permeability) as a complex quantity comes from the fact that, in general, the polarization of a given material always respond with some time delay after the application of an external electric field. The response is, in general, dependent on frequency of the applied field. Therefore, one needs to specify not only the magnitude but also a phase difference in order to model the causal relationship. Since complex numbers are defined by a magnitude and a phase, they are mathematically suitable for that representation. The physical origin of this representation is therefore rooted on causality. Note that the permittivity may also require a tensorial representation, as for instance, in uniaxial crystals, which respond anisotropically.}. The permittivity is a complex quantity and therefore it can be expressed as\footnote{Similarly, the complex permeability ($\mu$) is written as $\hat\mu = \mu^{\prime} + {\rm i} \mu^{\prime\prime}$, and the following statements have a similar correspondence for the magnetization processes in the material.}:

\begin{equation}
\hat\epsilon = \epsilon^{\prime} + {\rm i} \epsilon^{\prime\prime},
\end{equation}

\noindent where Re$(\hat\epsilon) = \epsilon^{\prime}$ and Im$(\hat\epsilon) = \epsilon^{\prime\prime}$. The imaginary part of the permittivity represents the electrical dissipation related to the polarization process in the material. In fact, the polarization vector field does not follow the changes of the electric vector field instantaneously. In a sinusoidal steady state, the power dissipation density $\langle P_d \rangle$ is associated with the polarization through a time average involving of the electric $\vec{E}$ and the displacement flux density $\vec{D}$ vector fields, resulting in \cite{Hau89}: 

\begin{equation}
\langle P_d \rangle \equiv \langle  \vec{E} \cdot {\partial \vec{D} \over \partial t} \rangle 
= {\omega \over 2} | \hat{E} |^2 \epsilon^{\prime\prime},
\end{equation}

\noindent where the angular frequency is $\omega = 2 \pi f$ ($f$, frequency) and $\hat{E}$ is the representation of the vector field as a complex quantity. Therefore, in the literature one often defines the {\sl loss tangent} for the permittivity as:

\begin{equation}
\tan \delta ={\epsilon^{\prime\prime} \over \epsilon^{\prime}}. \label {LT}
\end{equation}

\noindent and similarly for the permeability.

Physically, the behavior of the imaginary and real constitutive parameters is connected by Kromers-Kronig (KK) relation \cite{Jac99}, an important aid towards finding reasonable extrapolations outside the data range. However, the KK relation has its limitations in practical terms \cite{Mil97}, as we will point out in the next sections, and hence new methods, relying partially on the KK relation (or independent of it) are usually necessary. Here we offer a relatively simple, self-consistent method that makes use of the KK relation and other assumptions. In Section 2 we describe our procedure in detail, in Section 3 we offer a study case which implements our method, and in Section 4 we outline some general remarks on our results.

\bigskip
\bigskip

\textbf{2. METHOD}

\bigskip
\bigskip

In this section we present in detail the procedure used to obtain an extrapolation for the complex permittivity, given a finite frequency dataset. For definiteness, we will fix our procedure to the complex permittivity, {\it but the same steps are applicable to the complex permeability, as the principles are essentially the same}. Hence, for now on we shall treat the permittivity only. 

\begin{center}
\begin{figure}[htbp]
\centering
\includegraphics[scale=0.5]{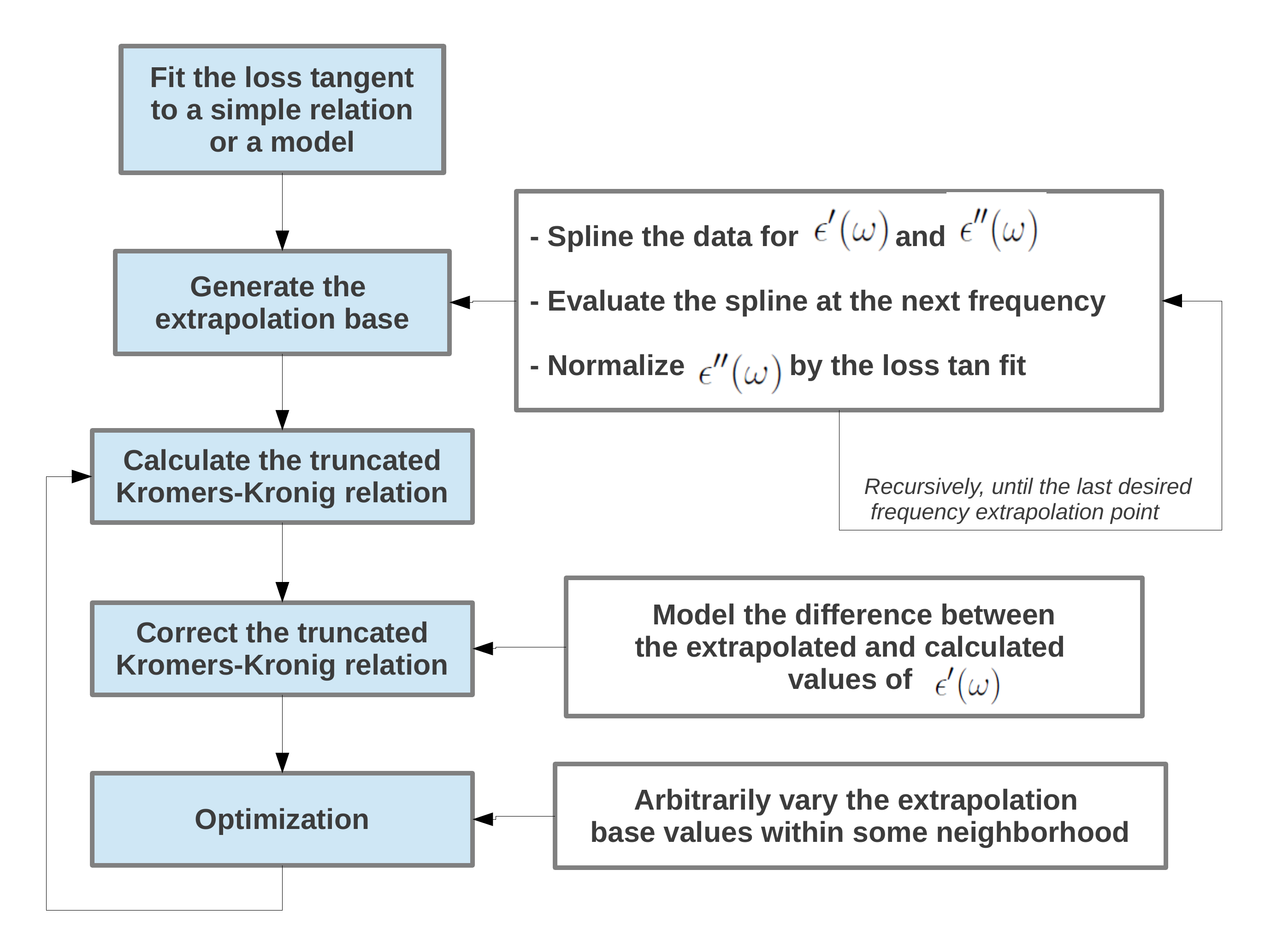}
\caption{Overall diagram of the method. The chained procedures at left represent the main logical steps. Diagrams at right indicate the chief computational or modeling operations for the given procedures.  \label{Diagram}}
\end{figure}
\end{center}

In Fig. \ref{Diagram}, we show a diagram with the main steps of the procedure in order to facilitate the understanding of the rationale behind the method, which is described in detail next.

\bigskip

\textbf{Step 1: To collect general hypotheses on the behavior of the constitutive parameters}

\bigskip

Given that the data is limited to a range of frequencies\footnote{Our extrapolation method was developed for data in the microwave range of approximately 2 GHz to 12 GHz, with the possibility of further extension up to about 25 GHz. However, it is experimentally known that possible deviations may occur above that value, depending on the propagation modes of  the electromagnetic wave in the waveguide. Our extrapolation procedure could still be applied in that case as, in general, the main characteristics of the attenuation behavior of the absorber are observed. However, further analysis is needed in those cases. Overall, our method is intrinsically robust for extrapolations up to about 100 GHz, if the hypotheses discussed in our paper are met.}, the behavior of the constitutive parameters outside the given range cannot be trivially inferred from the dataset without having some fundamental information on the material properties. Therefore, any extrapolation based on finite frequency data must be supported by some physical principles or hypotheses, and any further consideration or conclusion must strongly bear this caveat in mind. 

The first step therefore is to collect such hypotheses. The present problem offers a general principle to begin with: the fact that the behaviors (as a function of the radiation frequency) of the real part and the imaginary part of the permittivity are fundamentally related. In that regard, there are two basic inputs that are promptly available, namely: (i) the behavior of the loss tangent and its relation to the phenomenon of dimensional resonance, and (ii) the Kramers-Kronig relation \cite{Jac99}. Both offer a measure of the relation between the real and imaginary parts of the constitutive parameters. The first one is simply a result of the definition of the loss tangent:

\begin{equation}
 \epsilon^{\prime\prime}(\omega)  
= \epsilon^{\prime}(\omega) \tan \delta (\omega). \label {eppLT}
\end{equation}

\noindent where we explicitly wrote the dependence of the parameters on the angular frequency. A model for the behavior of the loss tangent as a function of frequency therefore establishes how the ratio between the imaginary and real parts of the permittivity vary as a function of frequency. One may start with the assumption that the limited information contained in the data offers a first approximation model to the loss tangent that is valid throughout the desired extrapolation frequency range, or one may already have some specific model in mind. In this paper we address the former situation, but the procedure is trivially valid for any model.

Extrapolating the behavior of the loss from finite frequency data is obviously delicate. For instance, in ferromagnetic materials, the loss tangent is expected to decrease approximately with the inverse of the frequency, except if large dielectric and magnetic losses are present due to dimensional resonance. One may find that the data does indicate some decreasing of the loss tangent with frequency. However, if dimensional resonance is present, the loss tangent may be oscillatory  at some frequency range outside the dataset. 

The condition for dimensional resonance can be estimated according to the following condition (see Eq. (3.24) in \cite{Soo60}):

\begin{equation}
d(f,\epsilon,\mu) > {c \over 2 f} {1 \over \sqrt {|\mu| |\epsilon|}}, \label {DR}
\end{equation}

\noindent where $d$ is the minimum length\footnote{Or thickness, for a normally incident radiation.} that the sample should have in order to present dimensional resonance, and $c$ is the velocity of light\footnote{The permittivity data was obtained from a network vector analyzer that linearly samples the frequencies in terms of ``dex'' intervals. The user may perform an appropriate change of variables in that formula for an otherwise different sampling scheme.}. Notice that only the moduli of the constitutive parameters are needed for the formula, considered in some average sense in the material. For increasing frequencies, the minimum length decreases.  This condition should be considered as an overall indicator of the behavior of the loss tangent for the extarpolated frequency range. If dimensional resonance regime is indicated, one should consider results based on extrapolations with even greater precaution, if no further specific physical models for resonance are input to the problem.

A second relation that connects the behavior of the imaginary and real parameters is given by Kromers-Kronig (KK) equation \cite{Jac99}:

\begin{equation}
 \epsilon^{\prime\prime}(\omega)  = 
1 + {2 \over \pi} \mathcal{P} \int_{0}^{\infty} 
{ \bar\omega\epsilon^{\prime\prime}(\bar{\omega})\over \bar{\omega}^2 - \omega^2} 
{\rm d} \bar{\omega}, \label {epKK}
\end{equation}

\noindent where $\mathcal P$ refers to the principal part of the integral. In our proposed scheme, we assume that the KK equation is a sufficient model for the behavior of the real and imaginary parts of the constitutive parameters of the material sample. In other words, we assume, that those parameters only depend on frequency and not on other parameters, such as position or direction within the material, or at least that any such dependence is not relevant to the problem at hand.

\bigskip

\textbf{Step 2: To model the hypotheses on the available data}

\bigskip

After considering the main hypotheses involved, the next step consists in modeling such hypotheses concretely. First, in our scheme, Eq. \ref{DR} is computed as a guide to what expect for the loss tangent in the frequency range where the data will be extrapolated. Second, a fitting curve for the behavior of the loss tangent in the range of the available data is obtained. It should suffice for our purposes to fit a simple curve in order to obtain a gross but discernible monotonic trend in frequency, if that is the case.  Third, natural cubic spline interpolations \cite{NR} to each of the imaginary and real parts of the permittivity are also performed.

One important problem with the KK relation (Eq. \ref{epKK}) in practical applications is the fact that {\sl the integral must be performed throught the frequency range} (from zero to infinity), and such information evidently is not available (it is exactly what we are searching for, namely,  the extrapolated data).  In order to circumvent this issue we use a {\sl truncated} version of the KK equation, where the limits of integration runs for the available data only, $\bar{\omega} = \{ \omega_{i} ..., \omega_{n} \}$, instead of the whole frequency range (from zero to infinity). It is clear that such a truncation is not guaranteed to give correct or accurate results. In order to address this issue, a crucial step in our scheme is to verify how far the truncated KK equation can be {\sl globally corrected} for the fact that only a much smaller range in the integration limits is used. By this we mean that a reasonable correction term to the KK equation, as a function frequency, up to some established order, $\mathcal{O}(n)$, should be searched for. By ``reasonable", we mean within some acceptable relative percentual error. In the next step, we clarify this issue and offer a possible correction procedure. 

Another problem is the numerical treatment of the singularity at $\omega = \bar{\omega}$. There are a few standard numerical treatments available to solve those problems. Here we adopt the scheme given in \cite{Ama95}, where the following singular integral is written as:

\begin{eqnarray}
\mathcal{I} & = & \mathcal{P} \int_a^b {g(\bar\omega)\over \bar\omega^2 - \omega^2} {\rm d} \bar\omega \nonumber \\
~ & = & \ln | \bar\omega^2 - \omega^2 | g(\bar\omega) \Big |_a^b -
 \left [  \left ( \bar\omega^2 - \omega^2 \right ) \ln | \bar\omega^2 - \omega^2 | -
\left ( \bar\omega^2 - \omega^2 \right )  \right ]
{{\rm d} g(\bar\omega)\over {\rm d} \bar\omega} \Bigg |_a^b + \nonumber \\
~ & ~ & + \int_a^b 
 \left [  \left ( \bar\omega^2 - \omega^2 \right ) \ln | \bar\omega^2 - \omega^2 | -
\left ( \bar\omega^2 - \omega^2 \right )  \right ]
{{\rm d}^2 g(\bar\omega)\over {\rm d} \bar\omega ^2} {\rm d} \bar \omega, \label{epKKn}
\end{eqnarray}

\noindent where $g(\bar\omega) \equiv  2\bar\omega \epsilon^{\prime\prime}(\bar\omega)$ in our problem. Therefore, $\mathcal{I}$ above is one half of that which appears in the KK relation (with a further $1/\pi$ factor), Eq. \ref{epKK}. Using that expression, with appropriate numerical treatment, we have a generally well-behaved integrand and the principal part of the KK integral can be well approximated.

\bigskip

\textbf{Step 3: To check the level of errors on the available data}

\bigskip

As previously mentioned, the truncated KK equation will result in an inaccurate determination of $\epsilon^{\prime}(\omega)$ because the given lack of complete knowledge of $\epsilon^{\prime\prime}(\bar{\omega})$ in the whole interval $\bar{\omega} = \{ 0, ..., \infty \}$. A simple scheme, which will be adopted here, is to first evaluate the truncated KK equation for the dataset (no extrapolation points). Then one computes the {\sl subtraction of the resulting truncated KK prediction to the actual data}. If the difference curve shows some relatively monotonic behavior with frequency, then this can be used as a global correction. For instance, it will be seen in the next section that in the worked example we are able to find an algebraic fit to the difference curve. One then recalculates the truncated KK equation plus the fitted difference curve as a correction. Subsequently one checks if this offers a globally accurate result, being acceptable at some prescribed level\footnote{Notice that such global correction would be trivially expected if the KK equation being used was some kind of approximation without an integral term, but not otherwise.}.

After evaluating the truncated KK equation plus correction within the frequency data range, $\bar{\omega} = \{ \omega_{i} ..., \omega_{n} \}$, in order to evaluate the robustness of the correction to the available data, this procedure is entirely repeated for some extrapolated dataset under examination. Notice that once a global correction to the truncated KK equation is found,  throughout the extrapolated region, any mismatch between the prediction of the truncated KK relation {\sl within the range of the available data} is, up to the correction order, a measure of how far the extrapolated dataset deviates from the ``true" solution (namely, that which would otherwise reasonably match the behavior of the data). We shall address this issue in more detail in the worked examples in the next section.

In the extrapolated region, we will also use the loss tangent model as a guide to the extrapolated parameters. Therefore it is interesting to evaluate the relative percentural error if one uses the information on the loss tangent fitting curve. One should therefore repeat the error estimation analysis (for the dataset only), as explained above, using however the right hand side of Eq. \ref{eppLT} in place of the $\epsilon^{\prime\prime}(\bar{\omega})$ in the truncated KK equation.

\bigskip

\textbf{Step 4: To extrapolate the data: direct procedures or inverse problem optimization techniques}

\bigskip

Direct procedures for performing the extrapolation are indicated in cases where there is either some model available for the material, or as probes to constrain the possible parameter space. The latter serves as input for an inverse problem optimization method. 

\bigskip

\textbf{A.  Direct procedures}

\bigskip

In this subsection  we present a simple extrapolation procedure. We recall that in step 2 a natural cubic spline has been performed on the data, and therefore the first and second derivatives of the splined fit at the last data point are available. One may use these parameters as a basis of recursive direct extrapolation. 

Consider the experimental data, given in intervals of frequency, $\Delta \bar{\omega}$. One obtains the next permittivity values, evaluated at the next frequency point ($\bar \omega + \Delta \bar{\omega}$), from the splined function and its derivatives.  This will establish an ``extrapolated'' loss tangent value for the obtained points, namely: 

\begin{equation}
\tan \delta_{\rm extrapol}(\bar{\omega}+\Delta \bar{\omega}) \equiv {\epsilon^{\prime\prime}_{\rm extrapol}(\bar{\omega}+\Delta \bar{\omega}) \over
\epsilon^{\prime}_{\rm extrapol}(\bar{\omega}+\Delta \bar{\omega})}.
\end{equation}

\noindent The value above is compared to that defined by the previously computed fit for the data (based on the form $f^{-2}$). That is, we establish the ratio:

\begin{equation}
\alpha \equiv {\tan \delta_{\rm extrapol}(\bar{\omega}+\Delta \bar{\omega})\over \tan \delta_{\rm fit,data}(\bar{\omega}+\Delta \bar{\omega})},
\end{equation}

\noindent and the extrapolated imaginary permittivity point is ``normalized'', according to:

\begin{equation}
\epsilon^{\prime\prime}_{\rm ext,norm}(\bar{\omega}+\Delta \bar{\omega}) \equiv {1 \over \alpha}\epsilon^{\prime\prime}_{\rm extrapol}(\bar{\omega}+\Delta \bar{\omega} ),
\end{equation}

\noindent leaving the extrapolated real permittivity unchanged. Then, one runs again the truncated KK relation for the data with the above values, recursively throughout the desired extrapolation frequency range.

Clearly, the truncated KK relation will give an inaccurate result, and a correction has to be made in the same way that it was done in the steps above for the known data points. Yet, {\sl it is expected that the corrected KK relation should present improved results in the small scale sense}, because the frequency range has been enlarged and therefore the integral term, more accurately computed.

Once the {\sl corrected}, truncated KK relation is computed, one should check how far its predictions deviate from the data. The deviations express, in a minor sense, the error in the correction, up to the specified order, and, in a major sense,  a measure of how far the extrapolated behavior deviates from the true solution of the problem.

One might wish to explore some models or other extrapolation attempts in order to check the trends towards improving the solution, that is, in obtaining a better final match between the corrected truncated KK equation and the dataset. This could lead to an acceptable extrapolation model, or not. Alternatively, such direct attempts (or models) may be used as constraints for an optimization method, which we outline in the next item.

\bigskip

\textbf{B. Optimization techniques}

\bigskip

It is not the purpose of this paper to evaluate in detail the performance of various optimization techniques to our specific problem. The reader is referred to, e.g., references \cite{Van07} and \cite{NR} for a ``traditional view'' of the available methods, or to, e.g., references \cite{Eib03} and \cite{Mic00} for more ``alternative'' methods. 

The point here is that our procedure can be implemented into a function that receives as input the permittivity values of a limited dataset, performs all the steps described above, and outputs some global measure of the acceptance of the extrapolation solution. We suggest that this measure be the following. Compute the area below the dataset real permittivity (splined) curve and that of the KK calculated, real permittivity (splined) curve, {\it within the dataset frequency range}. Take the difference between both areas, and the best extrapolation solution is that which minimizes this difference.

\bigskip
\bigskip

\textbf{3. CALCULATIONS AND RESULTS}

\bigskip
\bigskip

\textbf{Study case: extrapolation of the complex permittivity}

\bigskip

In this section we consider a study case using actual data obtained for a composite material, as described in detail in \cite{Pin13}. The data ranges from $f = \{  8 , ..., 12 \}$ GHz. The desired range of extrapolation is $f = \{  12, ... , 80 \}$ GHz. This example should clarify the procedure described previously.

\bigskip

\textbf{Step 1: Collecting general hypotheses on the behavior of the constitutive parameters}

\bigskip

\begin{center}
\begin{figure}[htbp]
\centering
\includegraphics[scale=0.47]{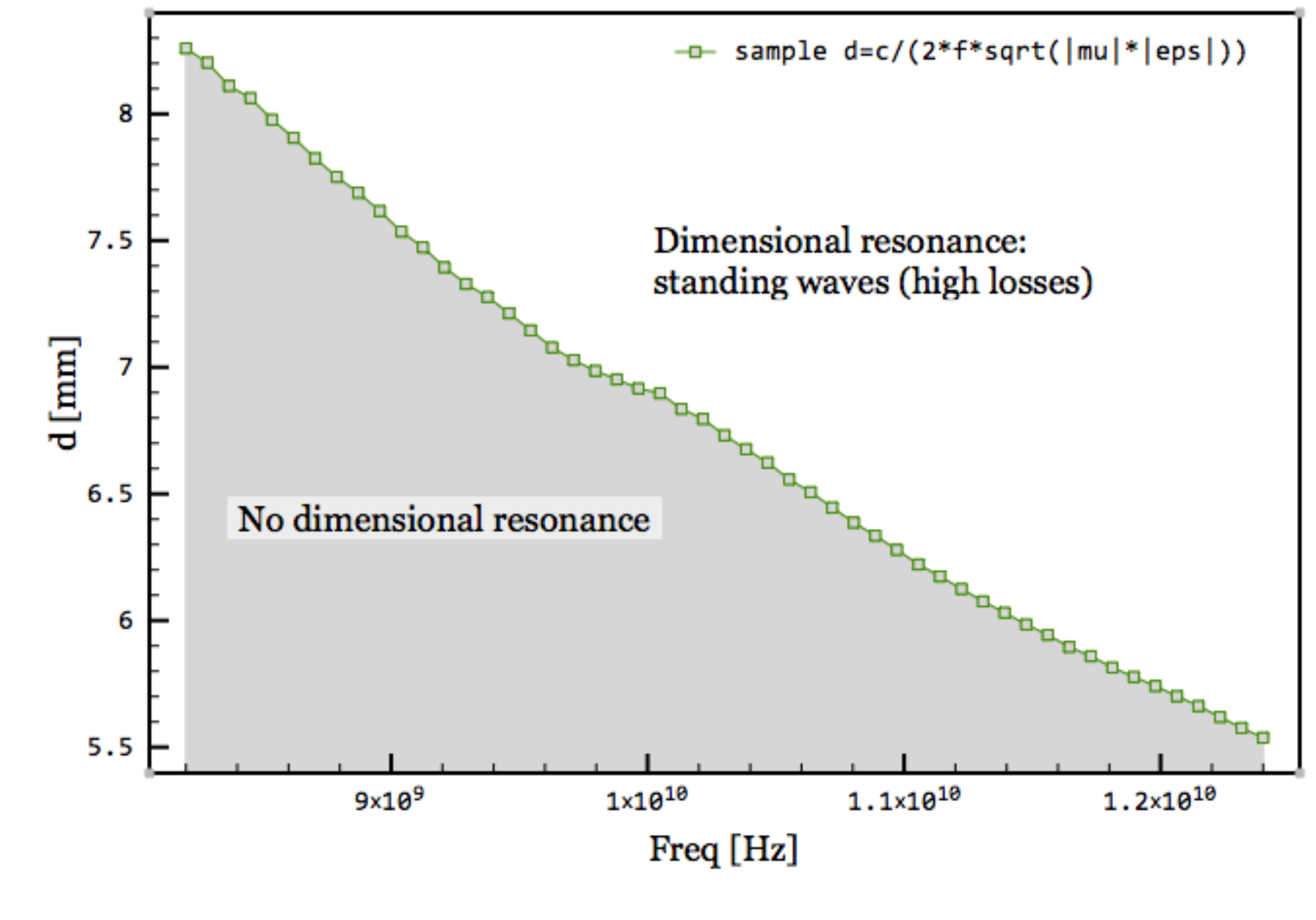}
\caption{ Dimensional resonance condition curve for the sample. \label{DimR}}
\end{figure}
\end{center}

We calculate the condition for dimensional resonance for the sample. A normally incident electromagnetic wave was imposed on the material with a thickness of $9$ mm.  We compute the value of $d$ in Eq. \ref{DR} for this sample and obtain the condition of the presence of stationary waves in the material, leading to losses.  Fig. \ref{DimR} shows that the regions defining resonance and non-resonance are separated by an approximately linear frontier curve throughout the frequency data range. The thickness of the present sample lies on the resonance region, as well as when considering the desired extrapolation range. 

From the expectation delineated above, the loss tangent might present fluctuations, whose magnitude and frequency are considered unknown. We begin by assuming, in a first order approximation, that the loss tangent function tends to follow the observed behavior in the dataset  qualitatively throughout the extrapolation interval.  In other words, even if the loss tangent happens to oscillate, we assume that, roughly, it does so around some simple monotonic function of the frequency. That is, in a first approximation, fluctuations are not considered important.

\bigskip

\textbf{Step 2: Modeling the hypotheses on the available data}

\bigskip

\begin{center}
\begin{figure}[htbp]
\centering
\includegraphics[scale=0.47]{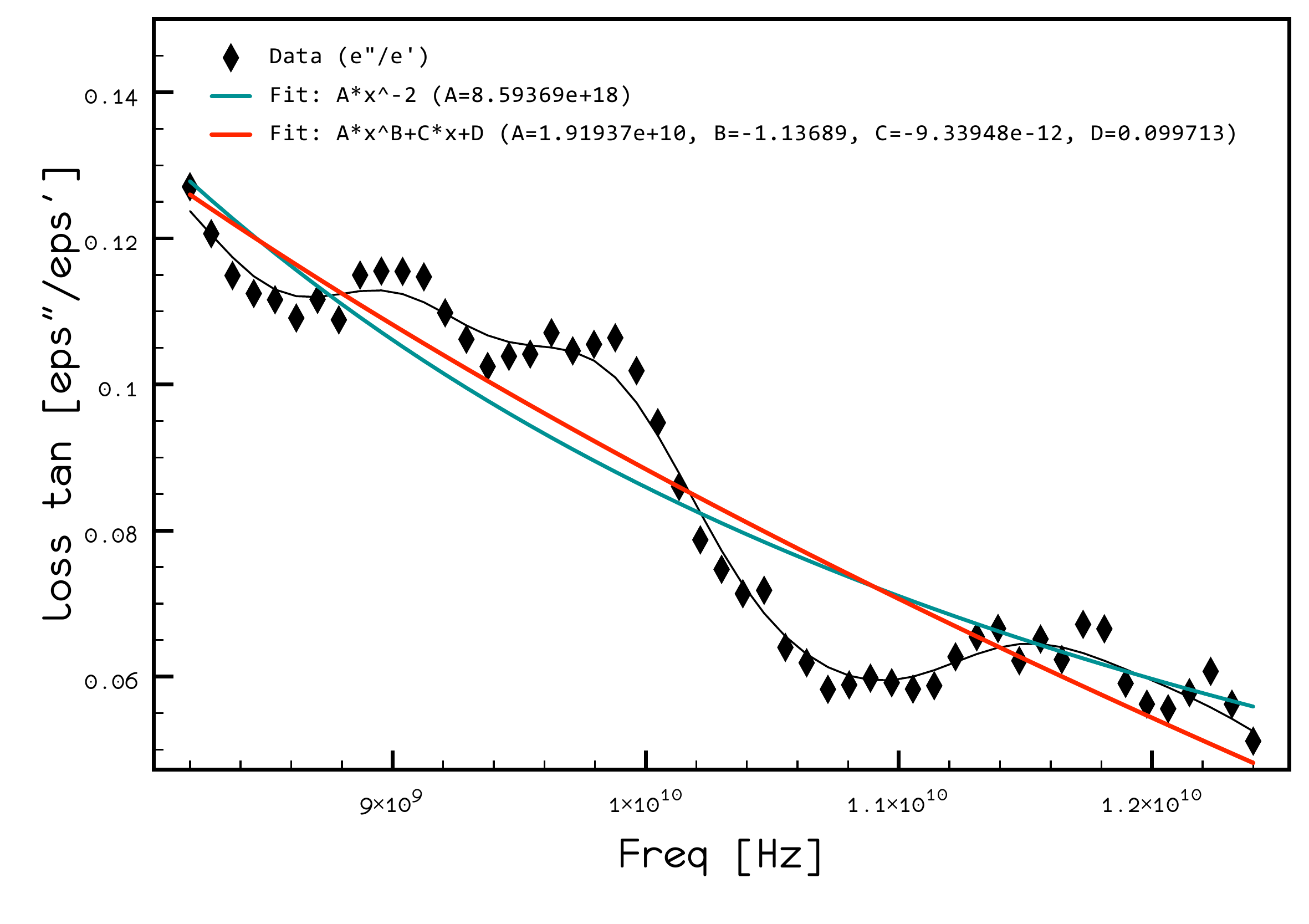}
\caption{Loss tangent fitting curves (splined) for an algebraic fit and for a simple $f^{-2}$ fit to the data. \label{Tan_Fits}}
\end{figure}
\end{center}

\begin{center}
\begin{figure}[htbp]
\centering
\includegraphics[scale=0.47]{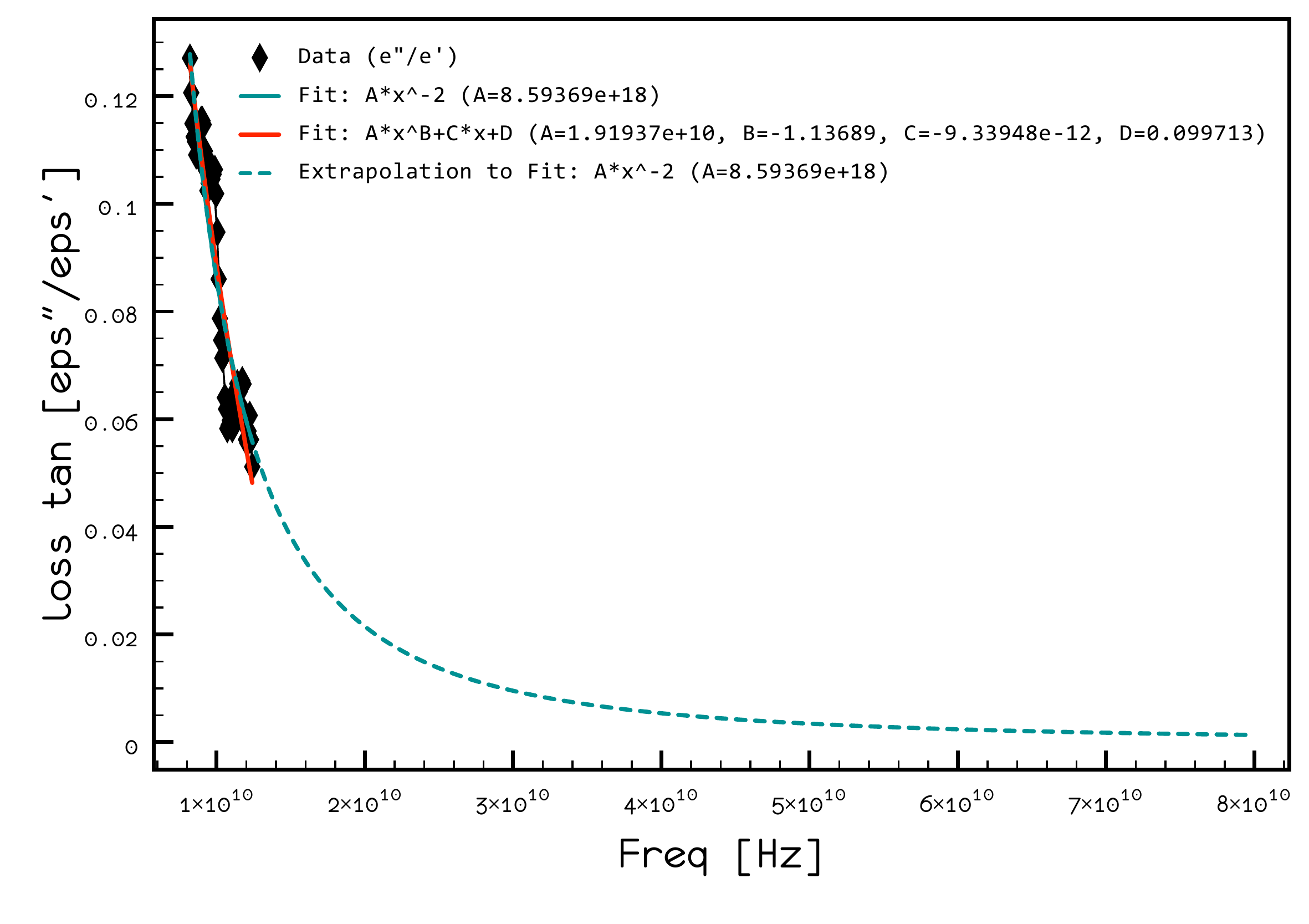}
\caption{The chosen extrapolated loss tangent curve (dashed curve). \label{Tan_Extrap}}
\end{figure}
\end{center}

We performed a cubic spline fit to the $\epsilon^{\prime}$ and $\epsilon^{\prime\prime}$ data, and also computed the corresponding loss tangent. In Fig. \ref{Tan_Fits}, we show the resulting (splined) loss tangent with an algebraic fit and a simple $f^{-2}$ fit. We choose the latter fit for two reasons: simplicity, and for the fact that it does not lead to a negative loss tangent in higher frequencies. The extrapolated loss tangent curve is shown in Fig. \ref{Tan_Extrap}.

Next, we compute the $\epsilon^{\prime}(\omega)$ function that the truncated KK relation would give based on the $\epsilon^{\prime\prime}(\bar{\omega})$ given by the data: (i) using direct data on $\epsilon^{\prime\prime}(\bar{\omega})$ itself into the integrand of Eq. \ref{epKK}; and (ii) using the loss tangent fit, $\sim f^{-2}$, described above, computed in the data frequency range (and also using the data on $\epsilon^{\prime}(\bar{\omega})$, see Eq. \ref{eppLT}) to input an ``approximated" $\epsilon^{\prime\prime}(\bar{\omega})$ for the integrand in Eq. \ref{epKK}.

\begin{center}
\begin{figure}[htbp]
\centering
\includegraphics[scale=0.5]{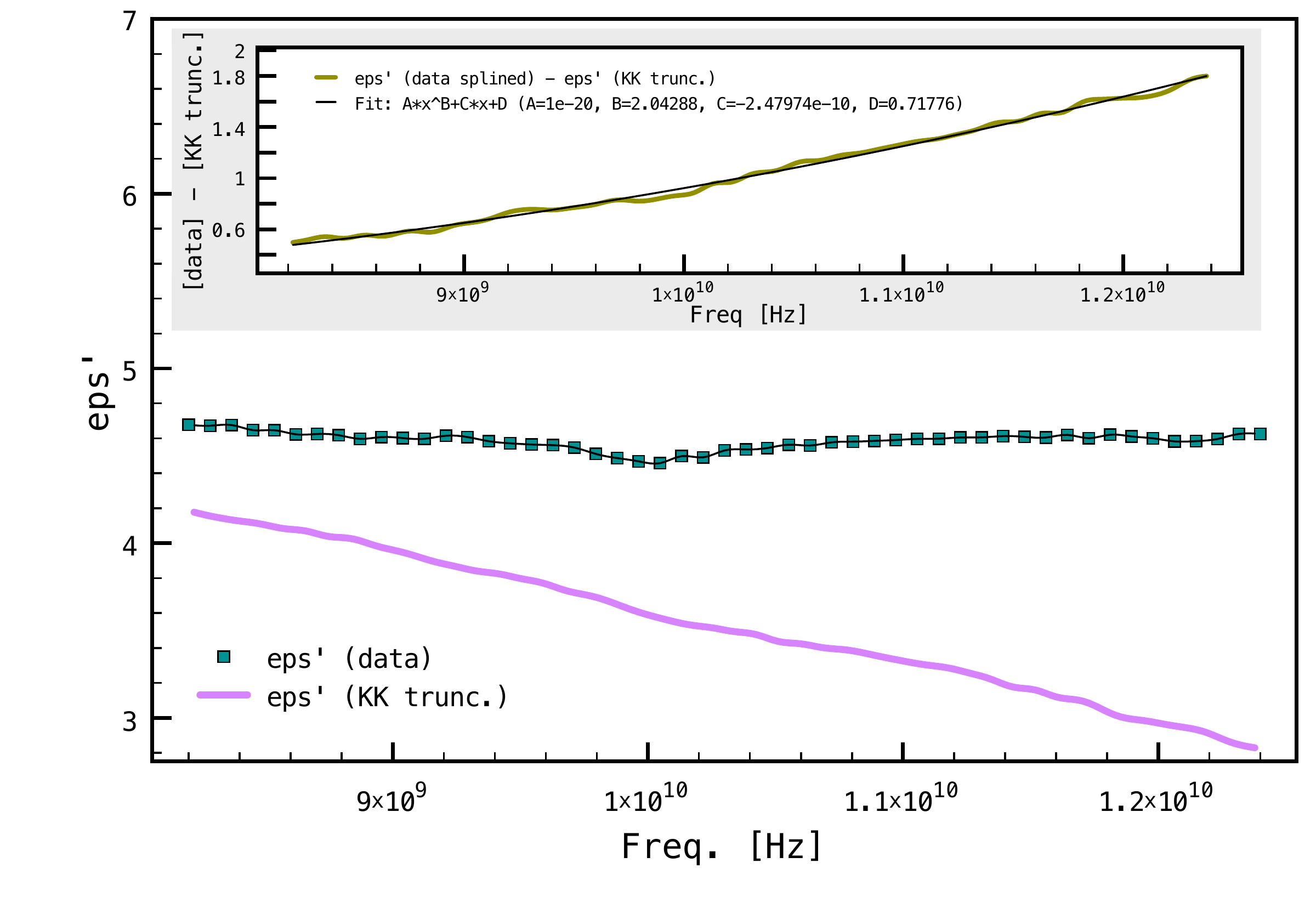}
\caption{Truncated KK result using direct data. \label{Epsilon_KK_Before_Fit_DATA}}
\end{figure}
\end{center}

\begin{center}
\begin{figure}[htbp]
\centering
\includegraphics[scale=0.5]{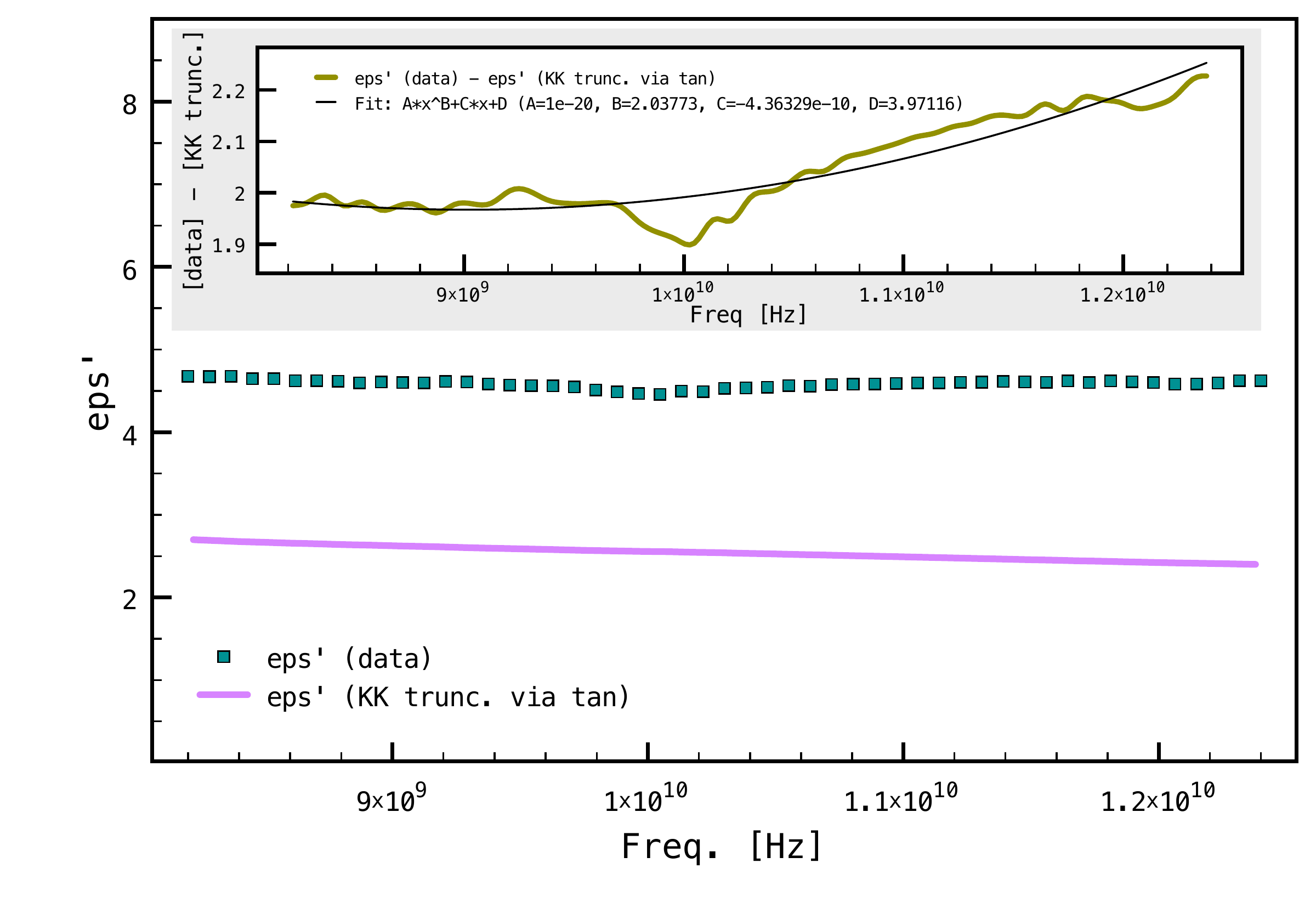}
\caption{Truncated KK result using the loss tangent fit, $\sim f^{-2}$. \label{Epsilon_KK_Before_Fit_LT}}
\end{figure}
\end{center}

In Figs. \ref{Epsilon_KK_Before_Fit_DATA} and \ref{Epsilon_KK_Before_Fit_LT}, we show the results for the cases (i) and (ii) above, respectively. It is clear that the truncated KK relation misses to reproduce the data on $\epsilon^{\prime}(\omega)$ in both cases, which is expected because the integration limits do not range throughout $0$ to $\infty$, as noted in the previous section. However, taking the difference between the splined data on $\epsilon^{\prime}(\omega)$ and the corresponding truncated KK estimate gives a monotonic curve (see insets on both figures). The difference curves are fitted then to an algebraic equation (a nearly second-degree polynomial), and the resulting fit is considered the correction to the truncated KK relation by simply adding it as a term to Eq. \ref{epKK}.

\bigskip

\textbf{Step 3: Checking the level of errors on the available data}

\bigskip

By using the ``difference curve''  fits described in the previous step as a correction to the truncated KK relation, we compute again its prediction for $\epsilon^{\prime}(\omega)$ for  both cases, (i) and (ii), as described previously.

\begin{center}
\begin{figure}[htbp]
\centering
\includegraphics[scale=0.47]{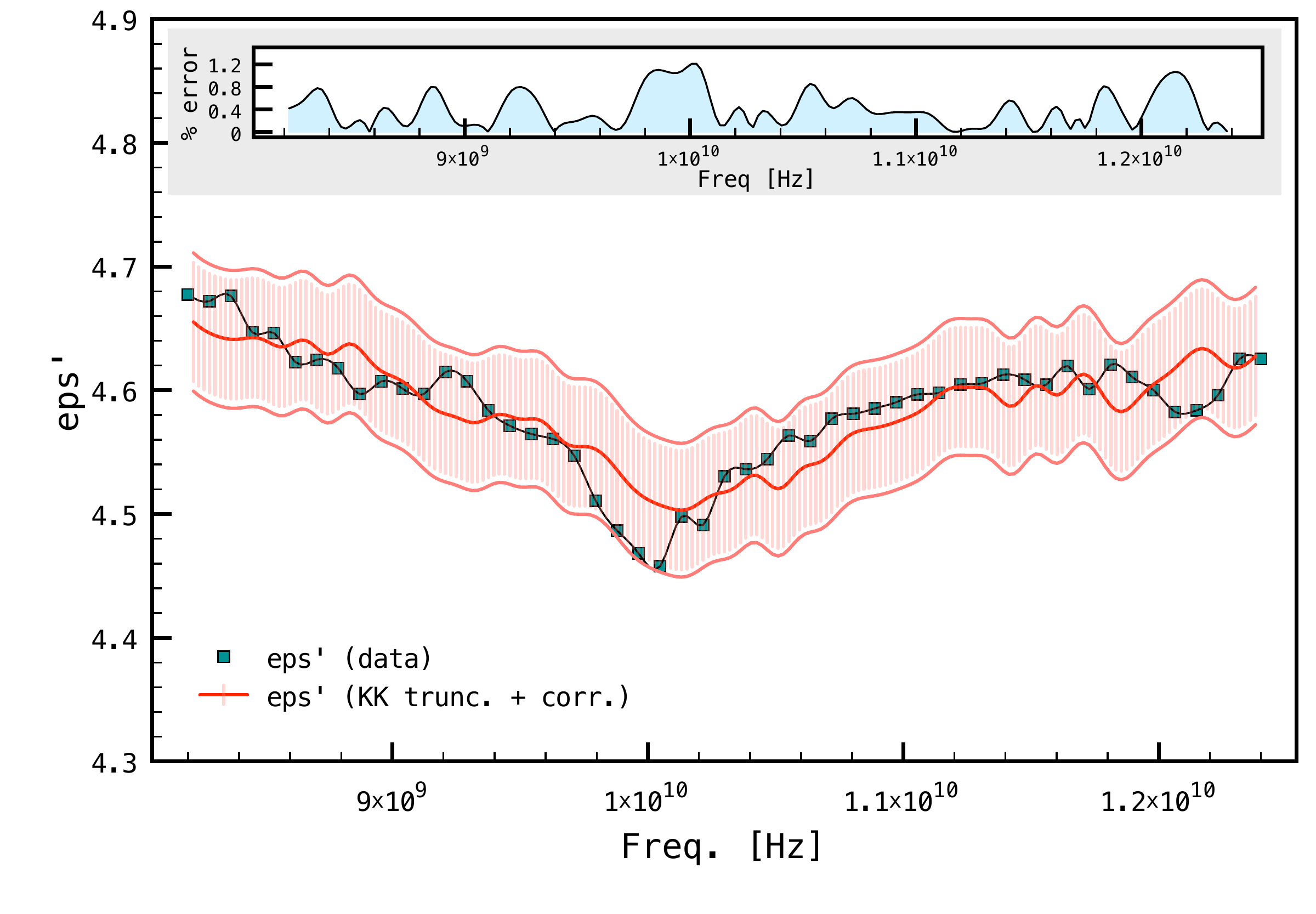}
\caption{Truncated KK result, with correction, using direct data. \label{Epsilon_KK_DATA}}
\end{figure}
\end{center}

\begin{center}
\begin{figure}[htbp]
\centering
\includegraphics[scale=0.47]{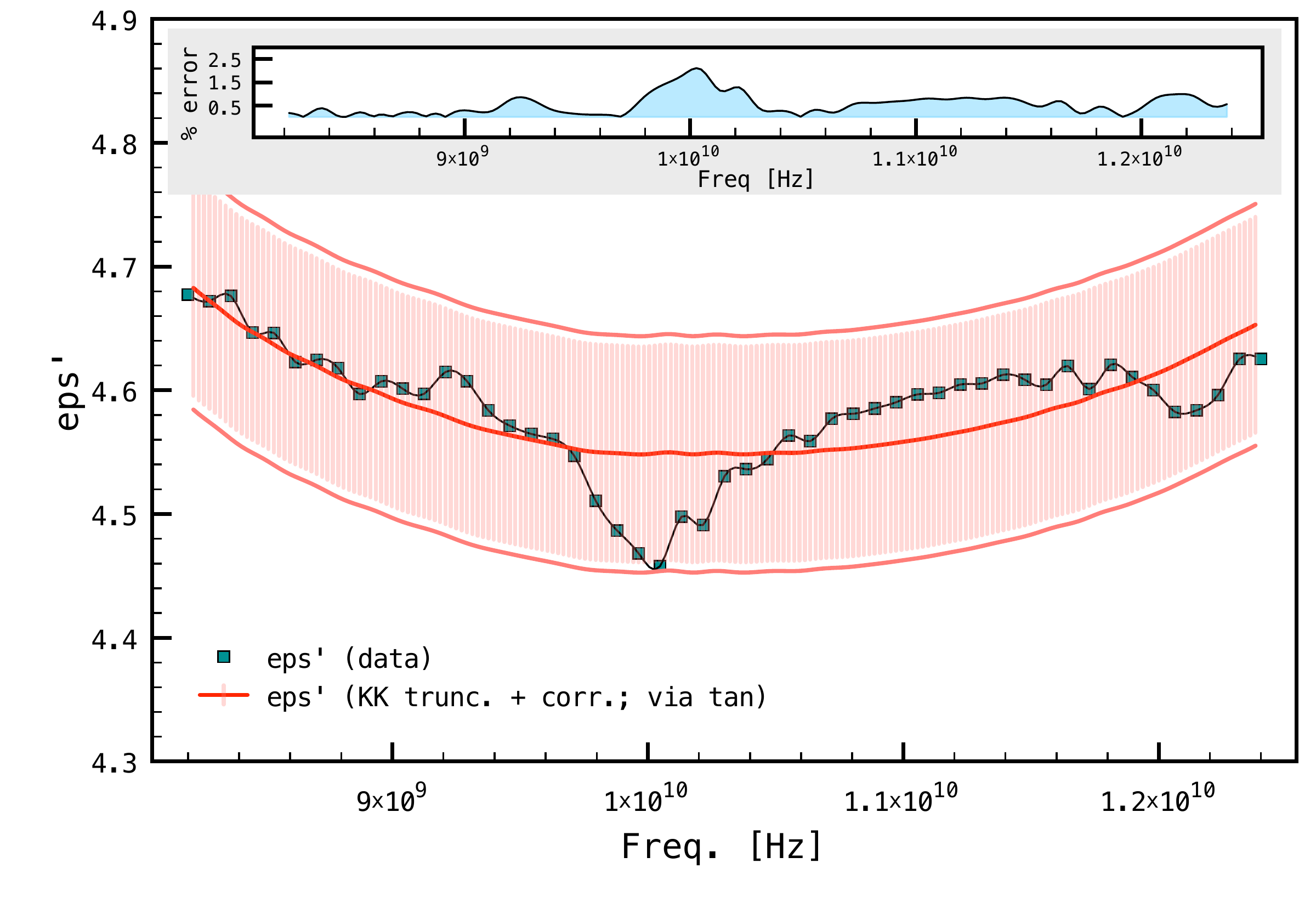}
\caption{Truncated KK result, with correction, using the loss tangent fit, $\sim f^{-2}$. \label{Epsilon_KK_LT}}
\end{figure}
\end{center}

Figs. \ref{Epsilon_KK_DATA} and \ref{Epsilon_KK_LT} show the results of the truncated KK relation with the corrections, made for cases (i) and (ii), respectively. We note that the truncated KK relation plus correction still fails to reproduce the ``fine-scale'' details of the data, but it does reproduce the overall trend of the data within an upper value of $\sim 1.2 \%$ and $\sim 2 \%$ of percentual error (relatively to the data), respectively, as shown in the insets of these figures. 

Such gross consistency would be expected for an overall correction, but more importantly is the fact that {\sl the correction itself is a monotonic function of the frequency} (see insets of Figs. \ref{Epsilon_KK_Before_Fit_DATA} and \ref{Epsilon_KK_Before_Fit_LT}). This is an important issue, as we again note that the KK correction depends on an integral involving the $\epsilon^{\prime\prime}(\bar\omega)$ {\sl throughout the frequency range}. In the truncated case, given that the integral is limited to a small range of frequencies, it is a nice feature that it errs by a factor that is a function itself of the frequency. It is not clear at this point whether this is a mathematically expected feature, or data dependent, or both.

In any case, these results show that the truncated KK relation, with ``global" correction, does offer, in the data frequency range, an ``order of magnitude'' estimate, within $\sim 2 \%$ of relative percentual error, to the real component of the permittivity, given either the imaginary component or the loss tangent data. Therefore, optimistically, a search for matching extrapolations is indicated, because the correction procedure to the truncated KK equation will probably not lead to increasing relative percentual errors (in fact, they are expected to improve on the inclusion of an increasing the frequency range). Any final deviations from the dataset will be, to second order in the present case, a measure of deviations from the true solution in the extrapolated range.

At this point we note that it is not trivial to set an error bar to the extrapolation, as it would come from the relative percentual error of the correction as well as from intrinsic deviations of the true solution, plus the hypotheses made. Therefore the indicated error bars in our present approach {\sl will always refer to the maximum percentual error within the data frequency range}.

\bigskip

\bigskip

\textbf{Step 4: Extrapolating the data}

\bigskip

\bigskip

\textbf{A. Direct procedures}

\bigskip

Now we adress the issue of probing the solution space in the extrapolated range by direct methods. We consider two cases:  extrapolating from some given model or recursively from the splinned data.

\begin{center}
\begin{figure}[htbp]
\centering
\includegraphics[scale=0.45]{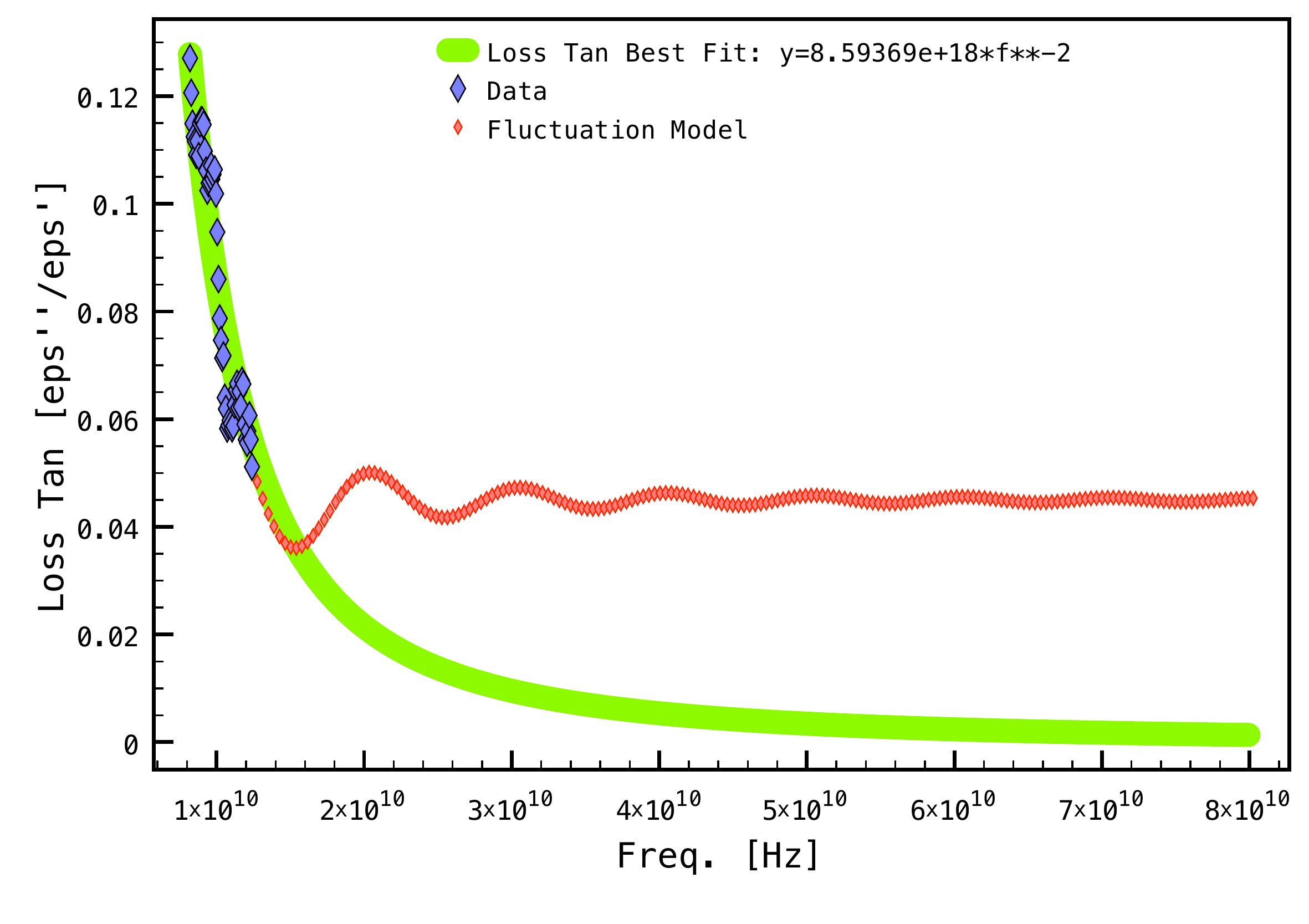}
\caption{An arbitrary fluctuation model for the loss tangent. Notice that it deviates from the loss tangent best fit model for the given data range. The direct extrapolation procedure is used in this model to study the intrinsic error in correction, as explained in the text. \label{Fluctuations}}
\end{figure}
\end{center}

\begin{center}
\begin{figure}[htbp]
\centering
\includegraphics[scale=0.38]{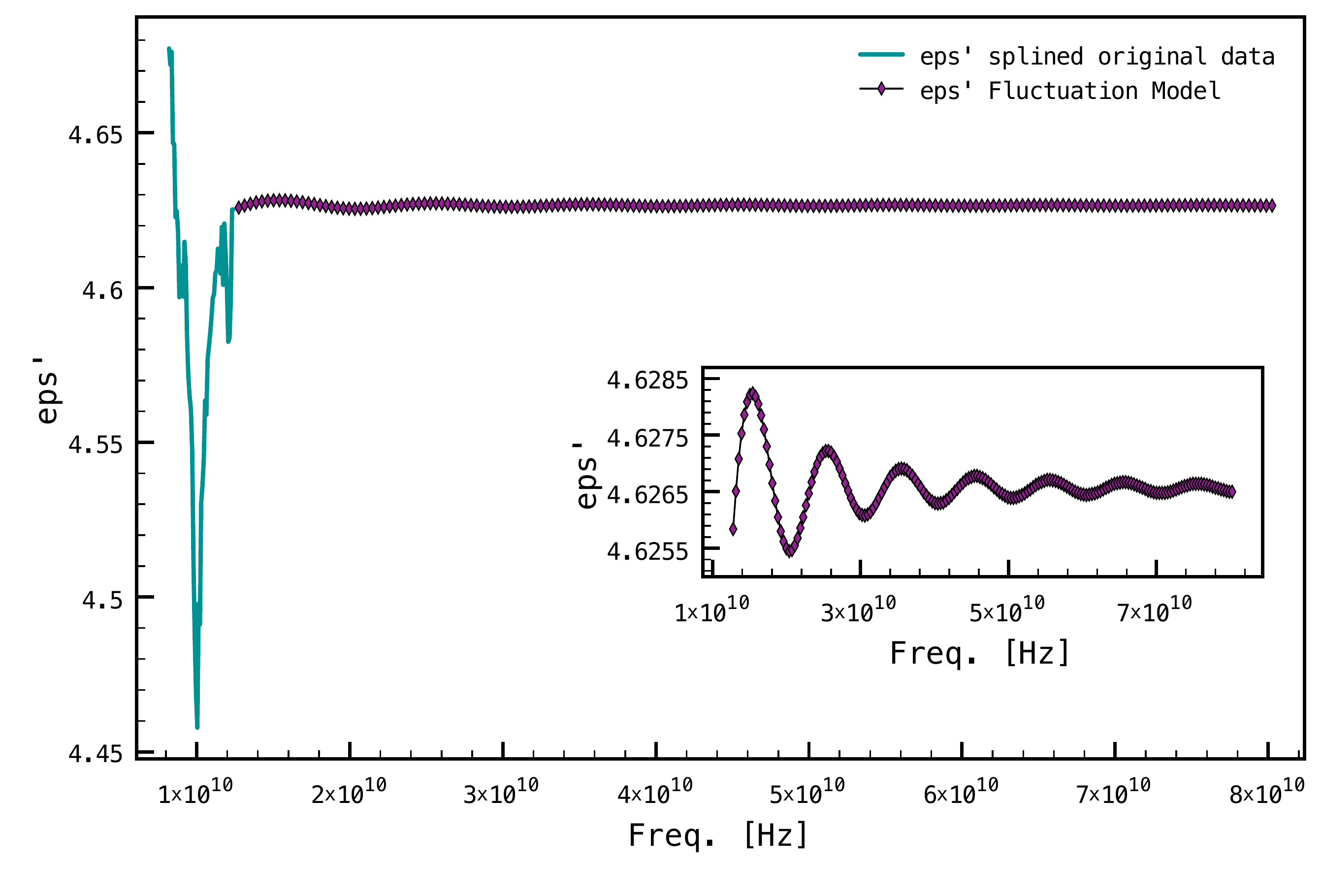}
\caption{The corresponding $\epsilon^{\prime}$ distribution for the arbitrary fluctuation model, obtained from the direct procedure in Sec. 2 (step 4A). The inset panel shows the model in more detail. \label{Fluc_ep}}
\end{figure}
\end{center}

\begin{center}
\begin{figure}[htbp]
\centering
\includegraphics[scale=0.38]{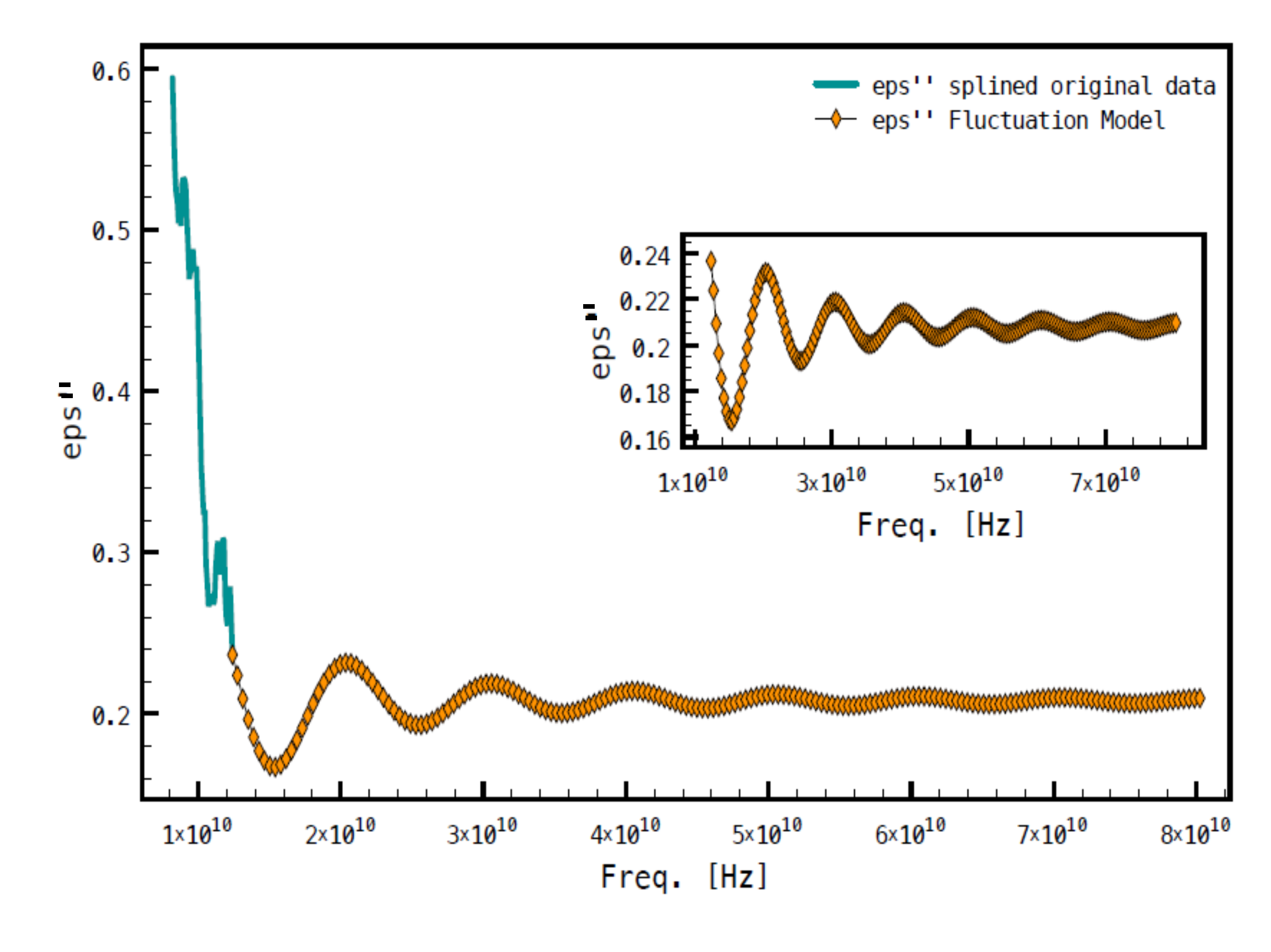}
\caption{The corresponding $\epsilon^{\prime\prime}$ distribution for the arbitrary fluctuation model, obtained from the direct procedure in Sec. 2 (step 4A). The inset panel shows the model in more detail. \label{Fluc_epp}}
\end{figure}
\end{center}

Suppose that one has a fluctuation model for the loss tangent, as shown in Fig. \ref{Fluctuations}. Here we use a completely arbitrary model just for illustrative purposes. The corresponding $\epsilon^{\prime}$ and $\epsilon^{\prime\prime}$ are shown in Figs. \ref{Fluc_ep} and \ref{Fluc_epp} with insets showing an amplified view of the level of fluctuations, respecitvely.

\begin{center}
\begin{figure}[htbp]
\centering
\includegraphics[scale=0.46]{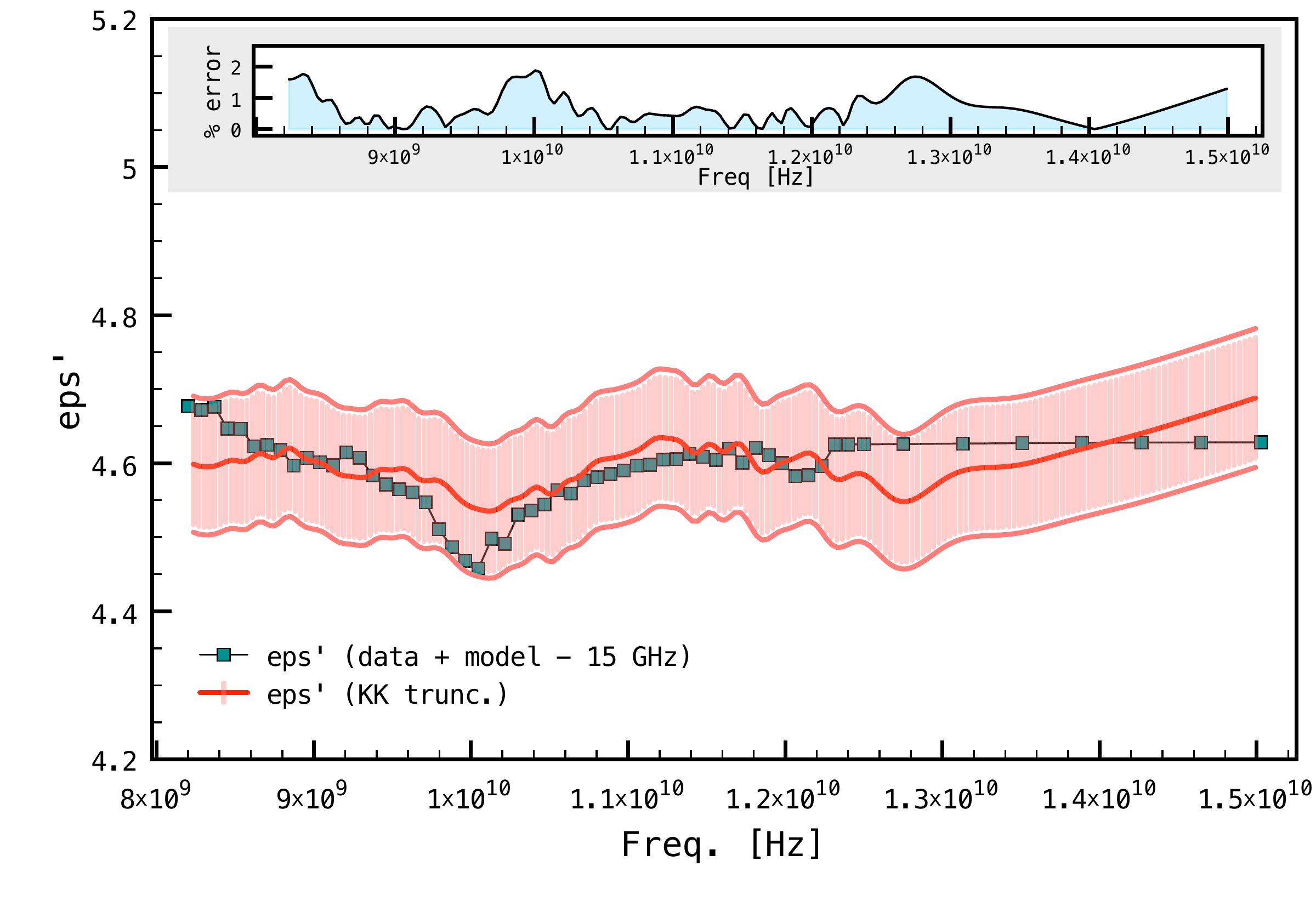}
\caption{Truncated KK result, with correction, for extrapolations up to $15$ GHz to the arbitrary model. \label{Epsilon_KK_15}}
\end{figure}
\end{center}

\begin{center}
\begin{figure}[htbp]
\centering
\includegraphics[scale=0.46]{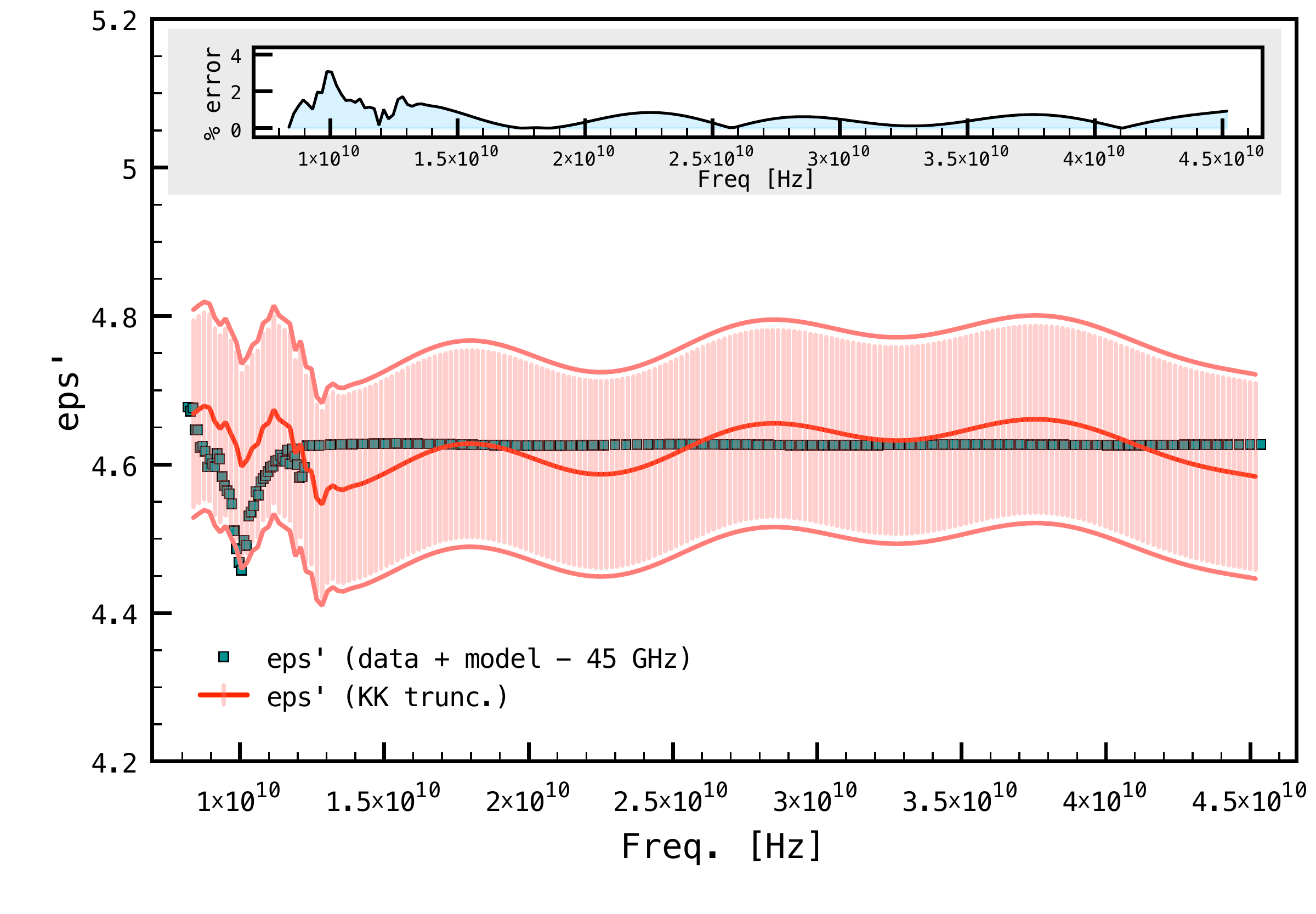}
\caption{Truncated KK result, with correction, for extrapolations up to $45$ GHz to the arbitrary model. \label{Epsilon_KK_45}}
\end{figure}
\end{center}

\begin{center}
\begin{figure}[htbp]
\centering
\includegraphics[scale=0.46]{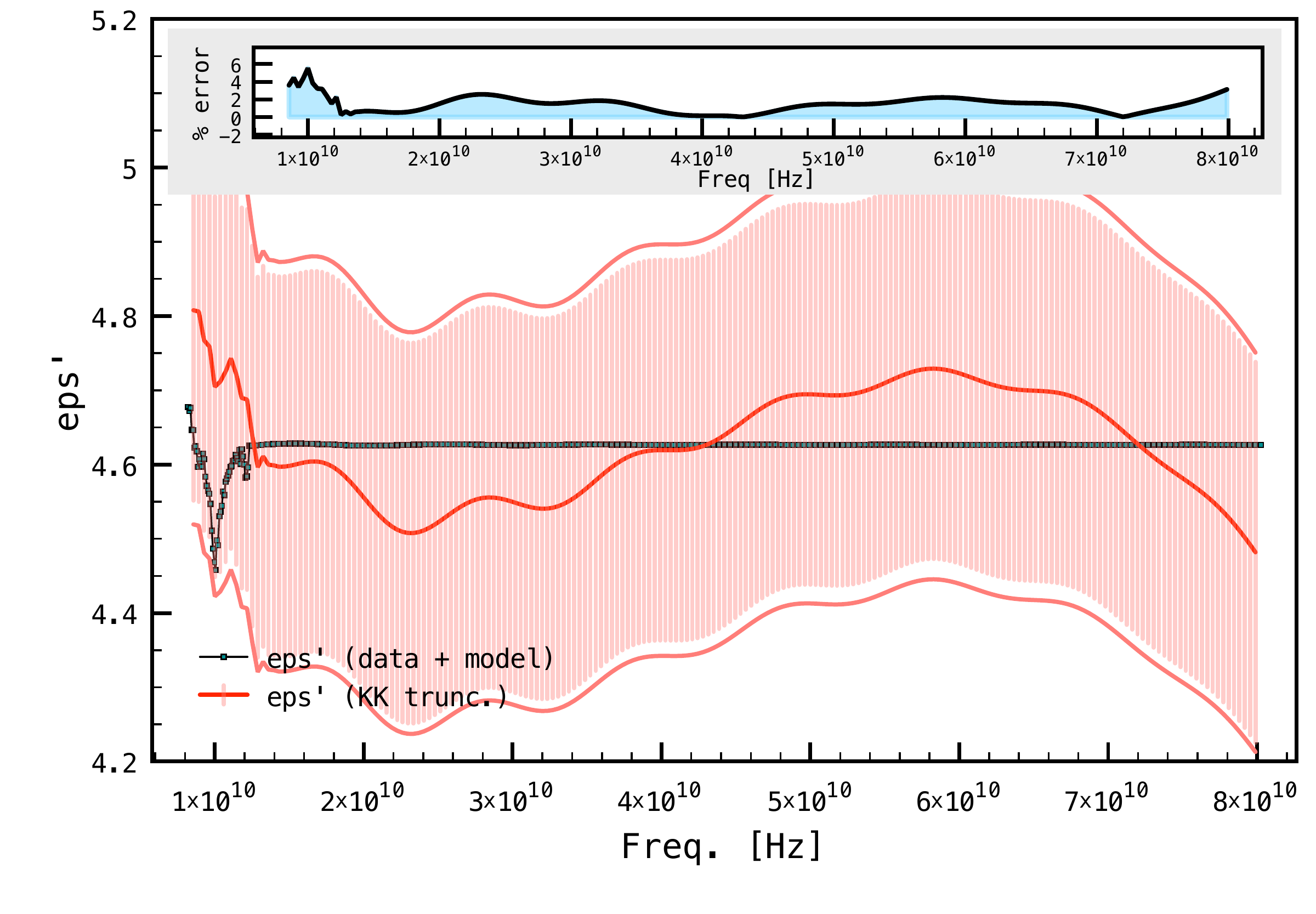}
\caption{Truncated KK result, with correction, for extrapolations up to $80$ GHz to the arbitrary model. \label{Epsilon_KK_80}}
\end{figure}
\end{center}

The model in question is analysed for extrapolations up to $15$, $45$ and $80$ GHz. The final results (with corrections) for the truncated KK relation are shown in Figs. \ref{Epsilon_KK_15}, \ref{Epsilon_KK_45}, and \ref{Epsilon_KK_80}. It is clear that the relative percentual error within the data range ($8$ to $12$ GHz) increases with the increasing extrapolating range. In the first figure, the relative error comes basically from the intrinsic error in correction. As we include more extrapolation points, the relative error includes more and more a true measure of the deviation from the real solution that otherwise would match the dataset more precisely.


\clearpage

Such direct probes can be attempted in an automatic fashion by feeding the computer code with relevant models. Then one checks by comparison those which better matches the dataset.

A second approach is that given in step 4.1 of previous section in which we spline the extrapolated point and compute the truncated KK relation recursively. The difference curve for the whole extrapolated range is obtained and given an algebraic fit. The difference curve fit is then used to correct to the truncated KK relation, which is re-calculated.

\begin{center}
\begin{figure}[htbp]
\centering
\includegraphics[scale=0.45]{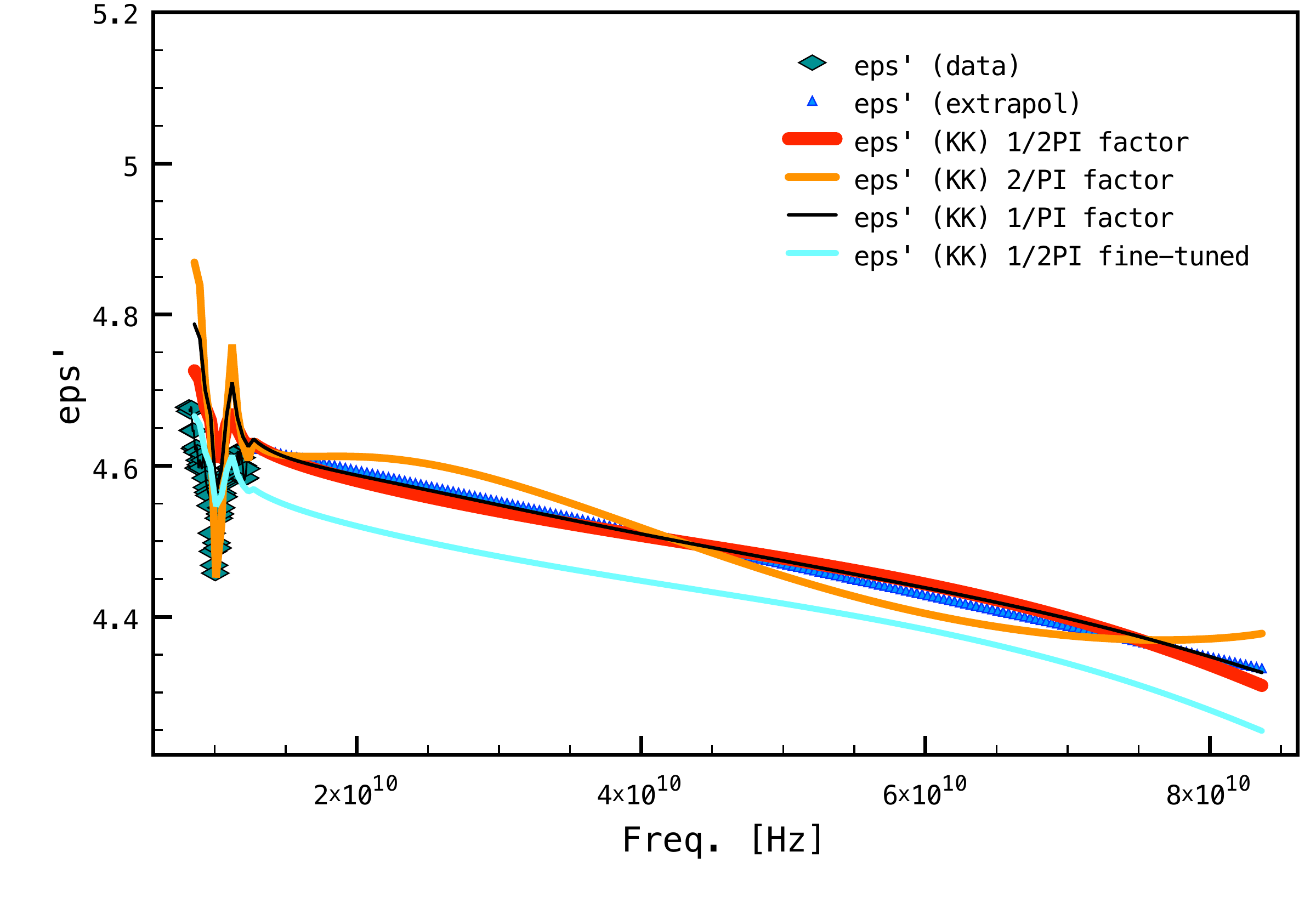}
\caption{Various extrapolation solutions based on the truncated KK relation, with correction, applied to the sample dataset, as explained in the text. \label{Epsilon_KK_EXT}}
\end{figure}
\end{center}

We have examined this procedure for various coefficients of the integral term of the truncated KK relation (originally set as $2/\pi$ in the original, ``non-truncated'' KK relation, Eq. \ref{epKK}). In one of the cases, we have interated once more the calculations by ``fine-tuning'' the correction, adding an overall small factor, determined by an ``eyeball'' examination of the $1/(2\pi)$ case. The results are shown in Fig. \ref{Epsilon_KK_EXT}.

\begin{center}
\begin{figure}[htbp]
\centering
\includegraphics[scale=0.45]{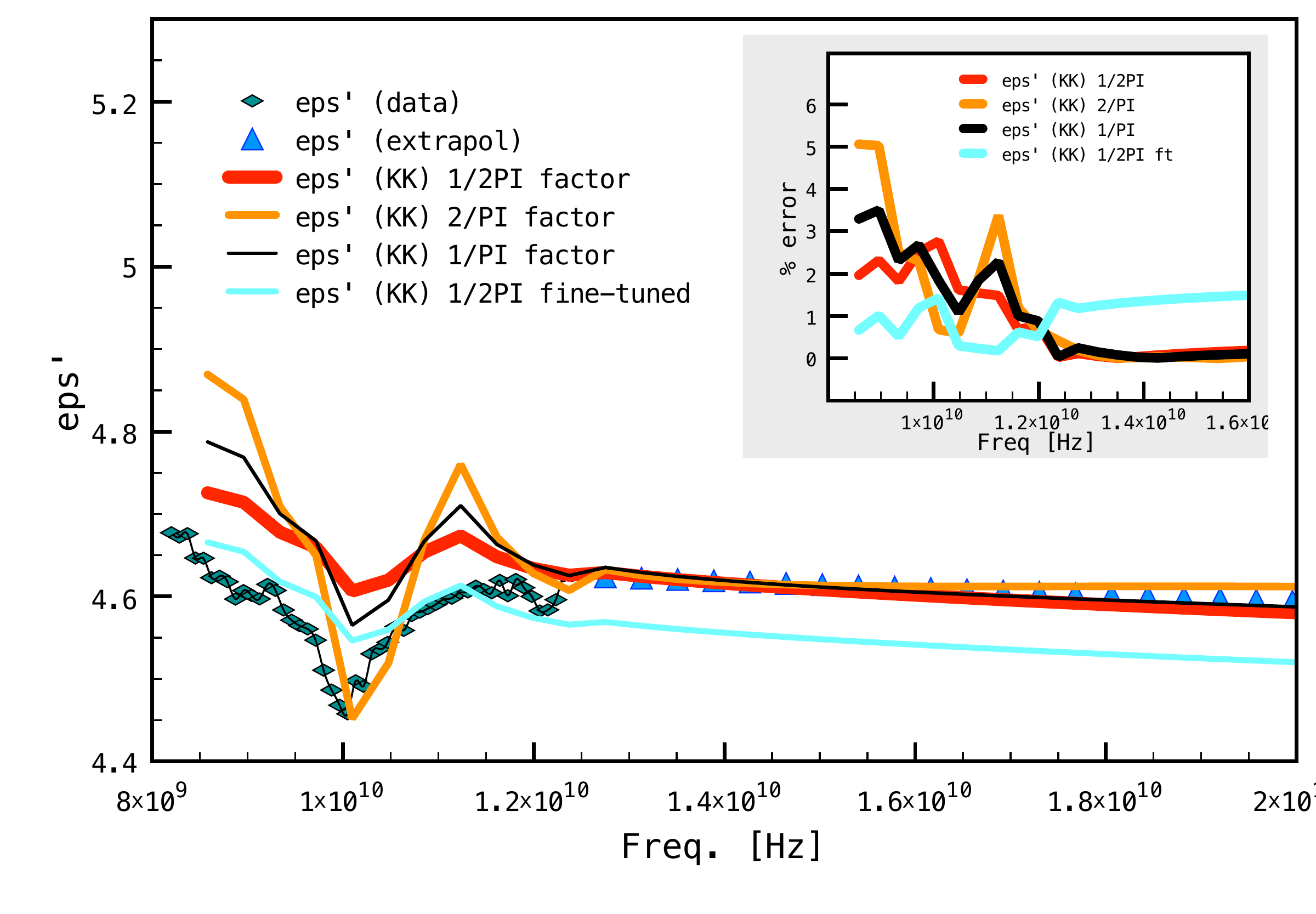}
\caption{A zoom out of the previous figure, in the range of the dataset, including part of the initial extrapolation range. \label{Epsilon_KK_Zoom_EXT}}
\end{figure}
\end{center}

Those extrapolations enable us to examine the prediction of the truncated KK relation, based on the extrapolated values within the data region. Fig. \ref{Epsilon_KK_Zoom_EXT} shows a zoom out of the previous figure and the relative percentual errors are shown in the inset. We observe that a larger coefficient in the integral term overestimates the fluctuations present in the data, with the exception of the dip present in the data, which is relatively well traced (see the $2/\pi$ case). On the other hand, the smaller coefficient, $1/(2\pi)$, traces well the fluctuations' amplitudes, except for the downward inflection in the data. As mentioned previously, we see that an increase in the frequency range of the integration limits of the truncated KK relation results in a better ``small-scale'' tracing of the fluctuations in the data. Such qualitative features can be taken into account in order to choose the extrapolation which better matches the underlying ``true''  solution.

\begin{center}
\begin{figure}[htbp]
\centering
\includegraphics[scale=0.45]{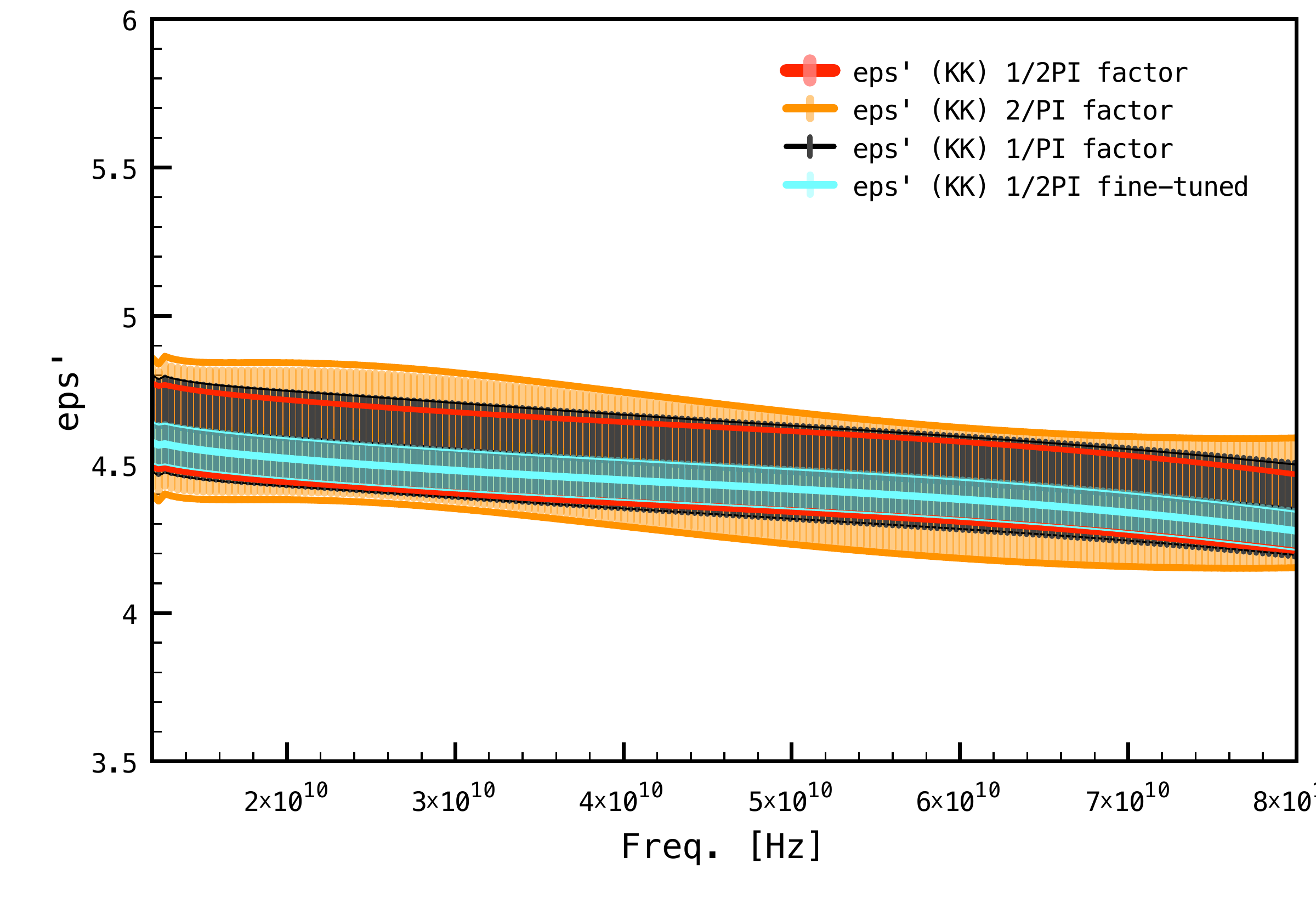}
\caption{Extrapolation results, in the extrapolated range only, including error bands, defined as the maximum relative percentual error in reproducing the original data, according to the results presented in Fig. \ref{Epsilon_KK_Zoom_EXT} \label{Epsilon_KK_final_extrapol}}
\end{figure}
\end{center}

As we mentioned, it is not trivial to quote the error bars of the extrapolated results. In Fig. \ref{Epsilon_KK_final_extrapol}, we include error bands for the extrapolations. The upper/lower limits of a given band along an extrapolation curve was set as the maximum relative percentual error in reproducing the original data, according to the results presented in the previous figure. Such error bands should not be interpreted literally as the errors involved in extrapolating the data. We can only state that values within error bands are collectively compatible with the data, and as such are a solution of the truncated KK relation, under the hypotheses stated.

\begin{center}
\begin{figure}[htbp]
\centering
\includegraphics[scale=0.45]{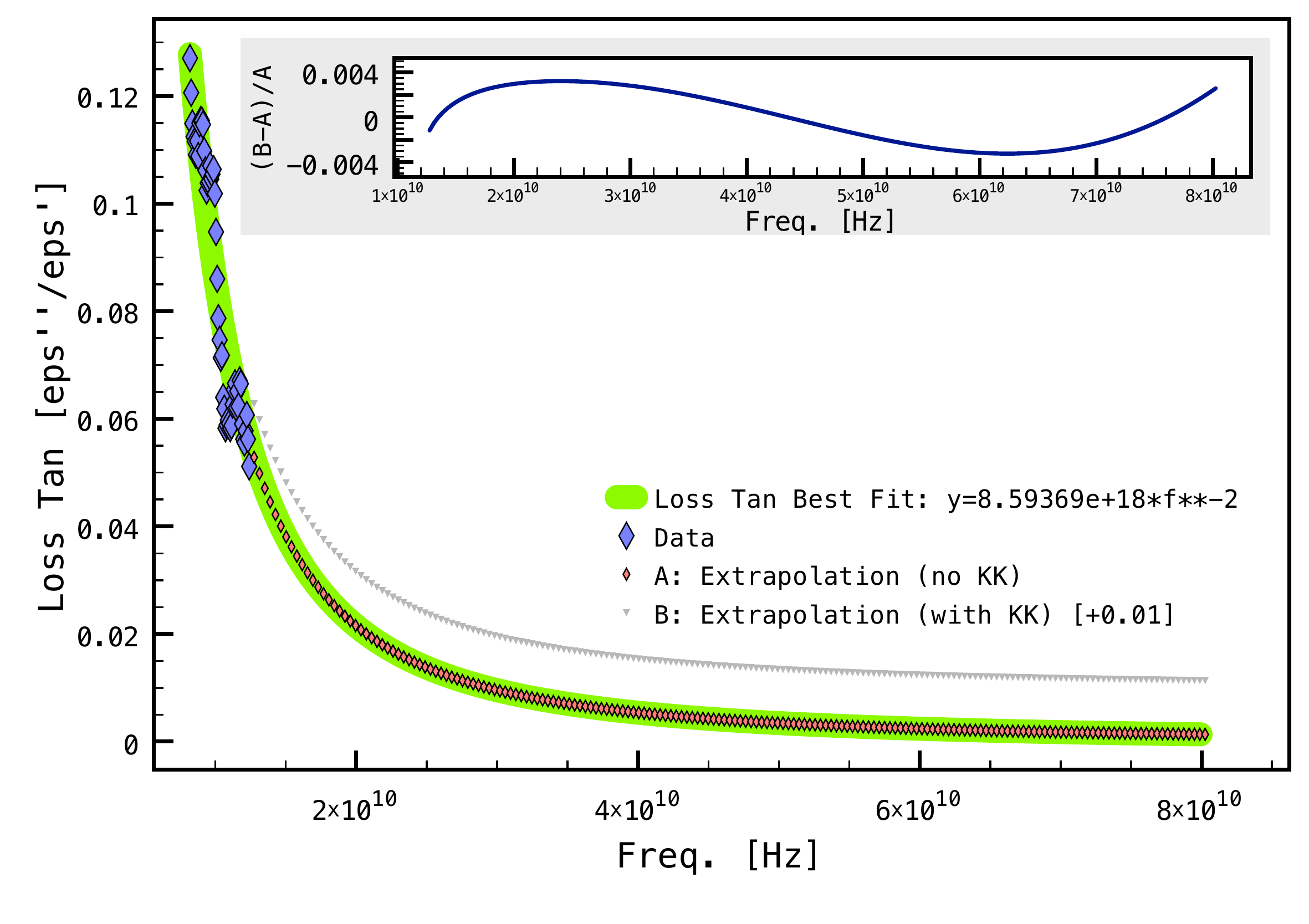}
\caption{The resulting loss tangent retrieved from the initially extrapolated data, and from one of  the extrapolated results (namely, ``fine-tuned'' model). The retrieved loss tangent will deviate a certain amount from the original fit. The relative difference between the retrieved loss tangent and the original one along the extrapolated frequence range is shown in the inset.
\label{Extrapolated_Tan}}
\end{figure}
\end{center}

In Fig. \ref{Extrapolated_Tan} we present the loss tangent for the initially extrapolated data (before calculating the $\epsilon^{\prime}$ values from the truncated KK relation), which was constrained to follow the $f^{-2}$ fit, but drawn from the splined data and its derivatives. We also show the loss tangent for one of  the extrapolated results.  The relative difference along the frequence range is shown in the inset. This view can also serve as a subsequent study model for the material properties.

\bigskip

\textbf{B. Optimization techniques}

\bigskip

\begin{center}
\begin{figure}[htbp]
\centering
\includegraphics[scale=0.75]{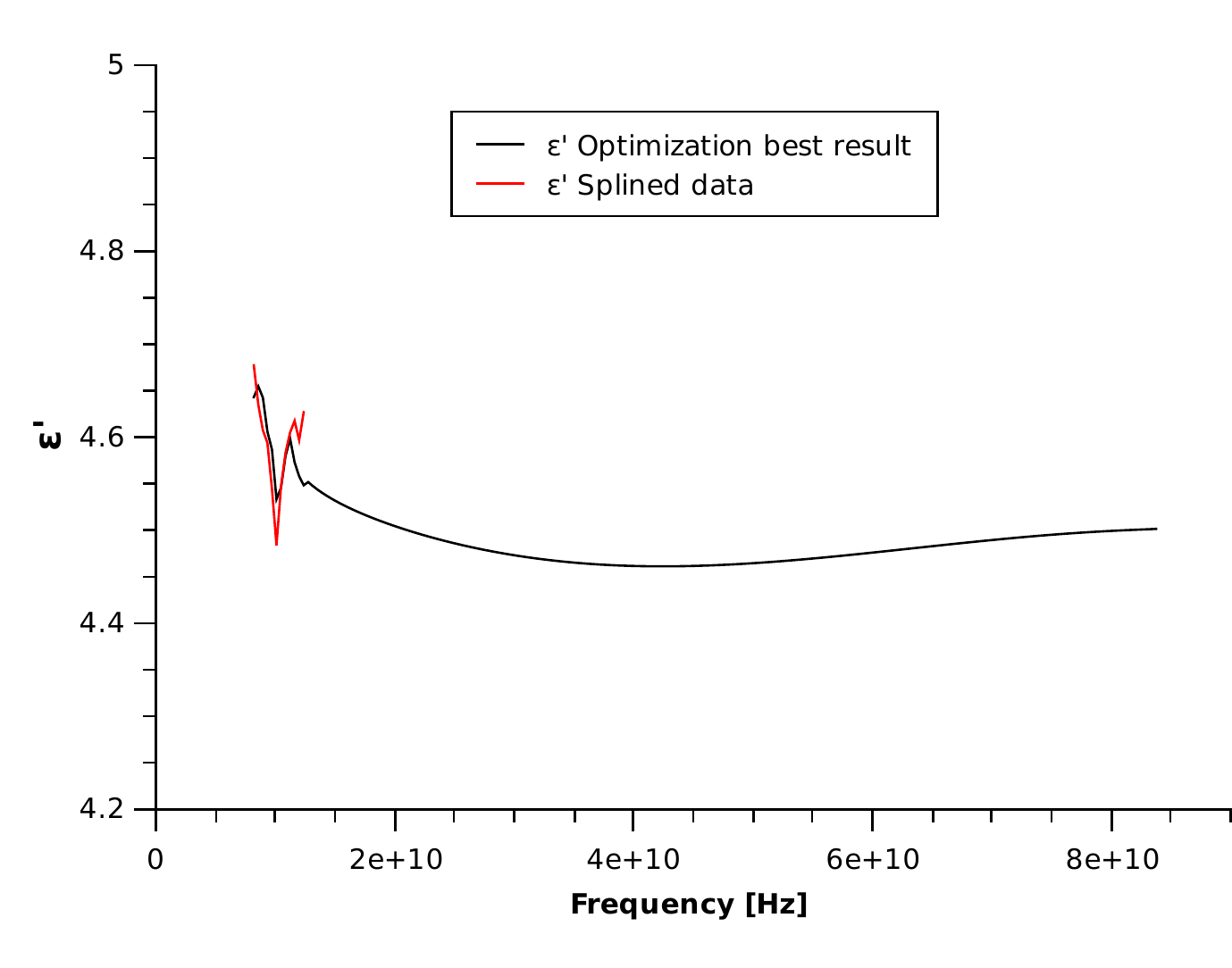}
\caption{The optimum extrapolation result. \label{Fig_Opt}}
\end{figure}
\end{center}

We implemented our whole procedure into one single function, as outlined in Section 2, step 4.2. In order to find the robustness of our procedure, we performed a simple Downhill Simplex Method \cite{NR} to that function, as it requires only evaluations on functions, and not their derivatives, and it is simpler to implement. After some tests, a tolerance parameter of $1.996$ was shown to be computationally adequate, resulting in $100$ evaluations in order to find the best solution. (Smaller tolerance values could be used to ``push'' the results into more optimized values, but it turned out to be computationally demanding for practical use). The minimum resulting difference in the areas between the two curves (see details in Section 2, step 4.2), however, was shown not to approach exactly zero (a perfect match). This is expected, as the method relies on a truncated KK evaluation and hence, it cannot match perfectly anyway. In any case, a minimum, optimized value for the extrapolated range can always be found, and it can be chosen as satisfactory. Our results are shown in Fig. \ref{Fig_Opt}.

\bigskip
\bigskip

\textbf{4. CONCLUSIONS}

\bigskip
\bigskip

The present work offers a procedure to extrapolate the complex permittivity (and permeability\footnote{Specific results for the permeability parameter were not shown, as the method is essentially the same.}) values of a material based on finite frequency data.  As any extrapolation, it is by its own nature a ``hazardous'' procedure (at least, as compared to interpolations, see, e.g., \cite{NR}). We have attempted to decrease the potential problems by considering some hypotheses, specially in what concerns the use of a ``truncated'' Kramers-Kronig relation, with an overall correction. 

We have evaluated in detail the level and nature of errors, and have found that, once an extrapolation base is fixed (e.g., under some hypotheses), the method is intrinsically robust to within a few percentual errors. However, it is important to note that we are not stating that the extrapolated results are correct to that level of errors. It is clear that experimental measurements should be performed in order to verify whether the hypotheses assumed initially stand against actual behavior of the data. Finally, we have tested the feasibility of the procedure for optimization techniques, by performing a simple test evaluation, which was shown to be adequate. We leave for future work a more thorough analysis of the better optimization strategy for the present problem.  

\bigskip
\bigskip

{\bf ACKNOWLEDGMENTS: We thank the financial support received from FAPESP (proc. \-2010/\ 09527-3 and CNPq (proc. 305478/ 2009-5). We thank the referees for useful suggestions and corrections.}

\end{document}